%% file: main.tex
\documentclass[conference]{IEEEtran}
\IEEEoverridecommandlockouts

\usepackage{cite}
\usepackage{amsmath,amssymb,amsfonts}
\usepackage{algorithmic}
\usepackage{graphicx}
\usepackage{textcomp}
\usepackage{soul}
\usepackage[table]{xcolor}

\def\BibTeX{{\rm B\kern-.05em{\sc i\kern-.025em b}\kern-.08em
    T\kern-.1667em\lower.7ex\hbox{E}\kern-.125emX}}

\RequirePackage{fix-cm}
\PassOptionsToPackage{table}{xcolor} 

\usepackage{enumitem}
\usepackage{amsmath,amsfonts}
\usepackage{algorithmic}
\usepackage{graphicx}
\usepackage{wrapfig}
\usepackage{textcomp}
\usepackage{listings}
\usepackage{tabularx}
\usepackage{pifont}
\usepackage{subcaption}
\usepackage{url}
\usepackage{balance}
\usepackage{tcolorbox}
\usepackage{wrapfig}
\usepackage{booktabs}
\usepackage[utf8]{inputenc}
\usepackage[normalem]{ulem}
\usepackage{graphicx}
\usepackage{pgfplots}
\usepackage{caption}
\usepackage{pgfplotstable}
\definecolor{codegreen}{rgb}{0,0.6,0}
\definecolor{codegray}{rgb}{0.5,0.5,0.5}
\definecolor{codepurple}{rgb}{0.58,0,0.82}
\definecolor{backcolour}{rgb}{0.95,0.95,0.92}
\definecolor{bblue}{HTML}{4F81BD}
\definecolor{rred}{HTML}{E11916}
\definecolor{ggreen}{HTML}{3FD72D}
\definecolor{ggreen1}{HTML}{9DEC9D}
\definecolor{ppurple}{HTML}{9F4C7C}
\definecolor{yyellow}{HTML}{FFC000}
\definecolor{yyellow1}{HTML}{FEE599}

\definecolor{debug}{HTML}{FFBABA}
\definecolor{info}{HTML}{FF5252}
\definecolor{warning}{HTML}{FF0000}
\definecolor{severe}{HTML}{A70000}

\definecolor{last-year}{HTML}{1E476C}
\definecolor{this-year}{HTML}{9FC5E8}

\lstdefinestyle{mystyle}{
  backgroundcolor=\color{backcolour},   commentstyle=\color{codegreen},
  keywordstyle=\color{magenta},
  numberstyle=\tiny\color{codegray},
  stringstyle=\color{codepurple},
  basicstyle=\ttfamily\footnotesize,
  breakatwhitespace=false,         
  breaklines=true,                 
  captionpos=b,                    
  keepspaces=true,                 
  numbers=left,                    
  numbersep=5pt,                  
  showspaces=false,                
  showstringspaces=false,
  showtabs=false,                  
  tabsize=1,   
  escapeinside={(*@}{@*)}
}

\newif\ifsubmission
\submissionfalse

\begin{document}

\title{\vspace{-1cm}{\normalsize\color{blue}\textbf{Accepted at the 2026 IEEE International Conference on Software Testing, Verification and Validation (ICST 2026)}}\\[2ex]
Assessing the Impact of Code Changes on the Fault Localizability of Large Language Models}

\author{
\IEEEauthorblockN{
Sabaat Haroon\textsuperscript{1},
Ahmad Faraz Khan\textsuperscript{1},
Ahmad Humayun\textsuperscript{1},
Waris Gill\textsuperscript{1},
Abdul Haddi Amjad\textsuperscript{1},\\
Ali R. Butt\textsuperscript{1},
Taha Khan\textsuperscript{2},
Muhammad Ali Gulzar\textsuperscript{1}
}
\IEEEauthorblockA{\textsuperscript{1}\textit{Virginia Tech, USA}}
\IEEEauthorblockA{\textsuperscript{2}\textit{Carnegie Mellon University, USA}}
}

\maketitle

\input{macros}

\input{section/abstract}


\input{section/introduction}
\input{section/motivation}

\input{section/research}
\input{section/Methodology}
\input{section/Evaluation}
\input{section/Discussion}

\input{section/related}
\input{section/conclusion}

\bibliographystyle{ieeetr}
\bibliography{main}

\end{document}

%% file: macros.tex
\ifsubmission
  \newcommand{\gulzar}[1]{}
  \newcommand{\faraz}[1]{}
  \newcommand{\ahmad}[1]{{\color{blue}{#1}}}
  \newcommand{\Sabaat}[1]{}
  \newcommand{\waris}[1]{}
  \newcommand{\haddi}[1]{{\color{olive}{#1}}}
\else
  \newcommand{\gulzar}[1]{{\color{red}{G: #1}}}
  \newcommand{\done}[1]{{\color{red}{DONE: #1}}}
  \newcommand{\faraz}[1]{{\color{orange}{F: #1}}}
  \newcommand{\ahmad}[1]{{\color{blue}{#1}}}
  \newcommand{\Sabaat}[1]{{\color{brown}{S: #1}}}
  \newcommand{\waris}[1]{{\color{cyan}{W: #1}}}
  \newcommand{\haddi}[1]{{\color{olive}{#1}}}
\fi

%% file: section/abstract.tex
\begin{abstract}
Generative Large Language Models (LLMs) are increasingly used in non-generative software maintenance tasks, such as fault localization (FL). Success in FL tasks depends on a model’s ability to reason about program semantics that are beyond surface-level syntactic and lexical features. However, widely used LLM benchmarks primarily evaluate code generation, which differs fundamentally from program semantic reasoning. Meanwhile, traditional fault localization benchmarks like Defect4J and BugsInPy are either not scalable or obsolete because their datasets have become part of LLM training data, leading to biased results. 
This paper presents the first {\em large-scale} empirical investigation into the robustness of LLMs' fault localizability. Inspired by mutation testing, we develop an end-to-end evaluation framework that addresses several limitations in current LLM evaluation, e.g.,  data contamination, scalability, automation, and extensibility.

Given real-world seed programs with specifications, we inject unseen faults and ask LLMs to localize them. We filter out underspecified programs, where correct fault localization is inherently ambiguous. For each program an LLM localizes successfully, we apply semantic-preserving mutations (SPMs) and rerun localization to assess the LLM's robustness and whether the LLM’s reasoning relies on syntactic cues rather than semantics.
We evaluate 10 state-of-the-art LLMs on $750{,}013$ fault-localization tasks sourced from over $1300$ Java and Python programs. We observe that SPMs cause an LLM to fail to localize the same fault it correctly localized earlier in 78\% of cases, and that LLMs' reasoning on the code found earlier in the context is noticeably better. These results suggest that  LLMs' code-reasoning is tied to code features irrelevant to semantics. We also identify code patterns that are challenging for LLMs to reason about. To the best of our knowledge, no prior work has evaluated the robustness of LLMs' code reasoning in fault localization at this scale. Overall, our findings motivate fundamental advances in how LLMs \emph{represent}, \emph{interpret}, and \emph{prioritize} code semantics to reason more deeply about program logic.


\end{abstract}

%% file: section/introduction.tex
\section{Introduction}


Large Language Models' (LLMs) growing adoption in software maintenance~\cite{HumanEval,Li2024LLMReasoning,Wang2025ReasoningLLMs} requires code \emph{reasoning} capabilities that differ fundamentally from code generation. In generation, models translate natural language descriptions into code, whereas software maintenance tasks demand understanding and operating on \emph{existing} code, often to answer questions such as fault localization or bug diagnosis. Consequently, while code generation is widely benchmarked~\cite{HumanEval,MBPP,codexglue}, limited benchmarking mechanisms exist for code reasoning tasks, and these studies typically suffer from data contamination, limited scalability, and insufficient rigor~\cite{deng-etal-2024-investigating,mem-bug,dong-etal-2024-generalization}. As a result, reported performance can be overly optimistic, and models may be overfitted to these benchmarks. 

LLMs are increasingly used for fault localization (FL) ~\cite{10.1145/3597503.3623342,SoapFL,CursorBugbot, sekiyama2024genai_troubleshooting_spark_glue}. Effective FL requires models to go beyond lexical or structural cues and reason about program \emph{semantics}, e.g., how state flows, how components interact, and how logic gives rise to observed failures. There is a critical need for systematic assessment of LLMs’ fault-localization capabilities for the future development of reliable coding models and autonomous agents. There are three key challenges in assessing LLMs' code reasoning for fault localization.

First, unlike traditional FL techniques that rely primarily on test suites, LLMs often require explicit specifications in addition to code to perform fault localization effectively. 
The absence of such specifications in many existing datasets, therefore, imposes additional constraints on evaluating LLMs.
Second, prior work~\cite{10.1145/3660771,10.1145/3597503.3623342,Nguyen2023FaultLocalization,SoapFL,10989036} frequently evaluate LLM-based fault localization on public benchmarks such as Defects4J~\cite{10.1145/2610384.2628055} and BugsInPy~\cite{10.1145/3368089.3417943}. Because these datasets are included in LLM pre-training corpora, the data is considered contaminated, leading to overly optimistic performance. Even with a newly constructed benchmark, LLMs undergo continuous training and are eventually exposed to public datasets\cite{deng-etal-2024-investigating}.
Third, evaluating LLMs’ code reasoning capabilities itself remains challenging due to the lack of standardized, scalable evaluation methodologies. 
Prior research has largely focused on human developers and relies on qualitative user studies, which do not readily translate to automated, large-scale evaluation of LLMs.
%

In this work,  we conduct the first large-scale empirical investigation of LLMs’ ability to reason about code for fault localization. To do so, we design an automated evaluation framework that dynamically generates controlled, previously unseen fault localization tasks to evaluate both accuracy and the robustness of LLMs’ code reasoning for fault localization.
Our insight is that if an LLM can correctly reason about a program’s semantics with respect to a specification, it should (a) identify deviations that change the intended behavior (i.e., faults) and (b) remain invariant to deviations that do not (semantic-preserving mutations). 
We address dataset contamination by dynamically injecting faults into real-world seed programs to create unseen fault localization tasks using classical mutation testing. 
For additional FL tasks, we apply semantic-preserving mutations (e.g., changes to comments and variable names, and the insertion of dead code) to faulty programs.
These transformations preserve the original specification while generating additional FL tasks, thereby reusing it and avoiding the need to write it manually. 
This approach is analogous to fuzzing, where diverse variants are generated through controlled transformations.

Given a set of source programs and their specifications, we first dynamically inject simple fault widely used in prior mutation testing research~\cite{10.1145/3526099}, including operator mutations (e.g., {\tt ==}, {\tt !=} , {\tt <}), conditional logic changes (e.g., branch inversion), constant and boundary modifications (e.g., off-by-one errors), and incorrect variable or return-value substitutions~\cite{bug_dissection,9463149}. 
%
We prompt each LLM with the faulty program and its specification and ask it to identify the injected fault’s line number. To avoid ambiguous FL tasks due to underspecified programs, we apply a counterexample-driven existential filter: we retain a (program, spec) instance if one or more LLM can localize the injected fault under that specification.
%

Since LLMs rely on attention mechanisms, they may overemphasize non-functional elements, such as comments or dead code, particularly when such patterns are frequent in the training data. 
%
We select only faulty programs correctly localized by the LLM and generate multiple semantic-preserving mutations (SPMs) that preserve program behavior and the injected fault to test the model’s fault localization robustness.
SPMs include identifier renaming, comment insertion or modification, formatting changes, and dead-code insertion, all of which are widely used in prior work~\cite{na-etal-2023-dip,RABIN2021106552}.
Each mutated program is paired with the exact same specification and provided to the LLM to re-identify the faulty line. 
%
By comparing fault localization responses across original, faulty, and mutated variants, we assess whether LLMs rely on true semantic cues or superficial code features.

We conduct our empirical investigation on state-of-the-art code datasets~\cite{iamtarun_python_2023,AhmedSSoliman_CodeSearchNet}, in which each program is accompanied by a specification describing its intended behavior. 
We start with 1,307 seed programs, comprising 637 Python and 670 Java programs. 
From these seeds, we inject 4 fault types and apply 6 semantic-preserving mutations, generating 750,013 unique faulty programs. 
The resulting dataset spans 245 million lines of code (LOC) and approximately 3.8 billion tokens.
We evaluate 10 state-of-the-art LLMs, spanning both closed-source commercial models (e.g., Open AI~\cite{OpenAI2023GPT4}, Claude~\cite{AnthropicClaude}) and open-source models (e.g., qwen~\cite{QwenModels}, phi4~\cite{Phi4TechnicalReport}).

Our investigation shows that even simple semantic-preserving mutations are sufficient to throw off LLMs’ reasoning capabilities. Across faulty programs that an LLM initially localizes correctly, applying SPMs causes LLMs to fail to localize the same fault in 78\% of cases. Further analysis reveals that misleading comments and dead code, both common in real-world programs, account for the majority of this robustness loss, with dead code alone reducing average accuracy to 20.38\%.  As expected, increasing the strength of SPMs produces a near-linear degradation trend: across all models, fault localization accuracy drops by 1.04\% per mutation step in Java and 1.93\% per step in Python, indicating compounding semantic interference.
Our evaluation framework is extensible and highly configurable. By controlling fault location, we find that 56\% of correctly localized faults appear within the first 25\% of program lines, compared to only 6\% in the final 25\%, highlighting strong positional bias. We also track LLMs' evolution in code reasoning for fault localization. Newer Claude and Gemini variants show very modest fault localizability gains (1–2\%), suggesting that recent model scaling and retraining have not substantially improved code reasoning required for FL.

To the best of our knowledge, this is the first empirical investigation of LLMs' code reasoning in fault localization tasks at this scale, supported by a fully automated, extensible evaluation framework.  Our findings indicate that LLMs’ code-reasoning in FL is highly nonlinear across the code surface and remains sensitive to code artifacts that are completely irrelevant to code semantics, thus motivating more fundamental advances in how LLMs \emph{represent}, \emph{interpret}, and \emph{prioritize} code to reason more deeply about code semantics.
%


\noindent\textbf{Data Availability.} The evaluation framework, datasets, and code generated and analyzed during the study are publicly available on Zenodo: \url{https://doi.org/10.5281/zenodo.18803969}

\begin{figure}[t]
\begin{lstlisting}[language=Python, numbers=left, numberstyle=\tiny, stepnumber=1, numbersep=3pt,
basicstyle=\ttfamily\scriptsize, keywordstyle=\color{blue}, commentstyle=\color{codegreen}, 
stringstyle=\color{red}, breaklines=true, frame=none, xleftmargin=1em]
def solveNQueens(n):
    def is_safe(board, row, col):
        for i in range(col):
            if board[row][i] == 1:
                return False

        for i, j in zip(range(row, -1, -1),
                        range(col, -1, -1)):
            if board[i][j] == 1:
                return False

        # Off-by-one fault: the loop stops one row too early
        for i, j in zip(range(row, (*@\textcolor{red}{n-1}@*), 1),
                        range(col, -1, -1)):
            if board[i][j] == 1:
                return False
        return True
    (*@\textcolor{gray}{<CODE REMOVED FOR BREVITY>}@*)
    def solveQueen(board, col, result):
    (*@\textcolor{gray}{<CODE REMOVED FOR BREVITY>}@*)
n = 4
solveNQueens(n)
\end{lstlisting}
    \caption{N-Queen program, $P_{F}$ ,  with an injected fault.}
    \label{fig:fault-motivating-example}
\end{figure}

\begin{figure*}[htbp]
\centering
\includegraphics[width=0.99\textwidth]{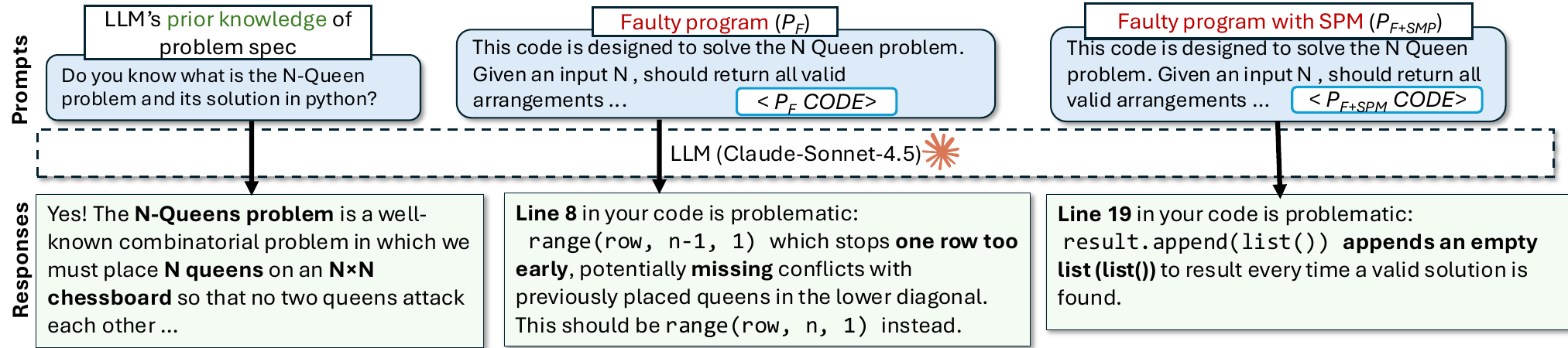}
\caption{Output of Claude's Sonnet 4.5 when asked about the N-Queen problem spec (first), asked for faulty line in the faulty code \textit{($P_F$)} (second), and asked for faulty line in the faulty code \textbf{\emph{with}} semantic-preserving mutations \textit{($P_{F+SPM}$)} (third) .}
\label{fig:LLM_debug_motiv}
\end{figure*}

%% file: section/motivation.tex
\section{Motivating Example}

We present a running example based on N-Queen problem taken from the Python dataset~\cite{iamtarun_python_2023}. 
To illustrate the limitations of LLMs’ code reasoning for fault localization, we inject a fault in this program and ask LLMs to identify the faulty line with and without semantic-preserving mutations.
%
We use Claude Sonnet 4.5 in this example. 
The N-Queen problem involves placing \(N\) queens on an \(N \times N\) chessboard so that no two queens threaten each other; that is, no two queens share the same row, column, or diagonal. 
This is a combinatorial optimization problem commonly solved using backtracking and constraint programming. 
We select the N-Queen problem because its algorithm is well-known, and its specification is unambiguous. 
We verify this by querying LLM about its specifications, which it accurately lists (Figure~\ref{fig:LLM_debug_motiv}-Left).  

For fault injection, we introduce an off-by-one fault in \texttt{is\_safe} function in line 13 as shown in red text in Figure~\ref{fig:fault-motivating-example}. 
The fault replaces \texttt{n} with \texttt{n-1}, which stops the exploration of the \emph{lower diagonal} of the board. 
The lower diagonal refers to the set of cells that lie diagonally downward (i.e., increasing row indices) and to the left (i.e., decreasing column indices) from a given position. 
Thus, a queen positioned along this diagonal can attack another queen located at the current position. We call this program $P_{F}$. 
We create another version of this faulty program by injecting three semantic-preserving mutations:  misleading function name, variable name, and comments, shown in Figure~\ref{fig:fault-motivating-example-spm}. 
Specifically, we rename the function name \texttt{solveNQueens(n)} to \texttt{howManyQueens(n)} and add a misleading comment (\emph{``This function checks how many queens are on the board.''}) and change the variable name \texttt{board} to \texttt{final\_result}. 
We call this program $P_{F+SPM}$. 
Next, we prepare two FL tasks for LLM, by first providing $P_{F}$ and then  $P_{F+SPM}$, the original N-queen specification, and ask the LLM to identify the faulty line of code in both. 
The prompt given to the LLM for both versions is:

\textit{``This code is designed to solve the N-Queen problem. Given an input \(N\), it should return all valid arrangements of \(N\) queens on an \(N \times N\) board such that no two queens attack each other. However, the code produces incorrect output. Can you identify the specific line of code responsible for the error? The program is attached below. {\tt <CODE>}"}

\begin{figure}[t]
    \centering
    \begin{lstlisting}[language=Python, numbers=left, numberstyle=\tiny, stepnumber=1, numbersep=3pt,
basicstyle=\ttfamily\scriptsize, keywordstyle=\color{blue}, commentstyle=\color{codegreen}, 
stringstyle=\color{red}, breaklines=true, frame=none, xleftmargin=1em]
(*@\textcolor{purple}{Misleading variable name.}@*)
def (*@\textcolor{blue}{howManyQueens}@*)(n):
    def is_safe(final_result, row, col):
        (*@\textcolor{purple}{Misleading comment.}@*)
        (*@\textcolor{blue}{\# This function checks how many queens}@*)
        (*@\textcolor{blue}{are on the board.}@*)
        for i in range(col):
            if final_result[row][i] == 1:
                return False

        for i, j in zip(range(row, (*@\textcolor{red}{n-1}@*), 1),
                        range(col, -1, -1)):
            if final_result[i][j] == 1:
                return False
        return True
    (*@\textcolor{gray}{<CODE REMOVED FOR BREVITY>}@*)
    def solveQueen(board, col, result):
        if col == n:
        result.append(list())
    (*@\textcolor{gray}{<CODE REMOVED FOR BREVITY>}@*)
(*@\textcolor{purple}{Misleading function name.}@*)
(*@\textcolor{blue}{howManyQueens}@*)(n)
\end{lstlisting}
    \caption{N-Queen program, $P_{F+SPM}$,  with an injected bug in Line 11 and Semantic Preserving Mutations.}
    \label{fig:fault-motivating-example-spm}
\end{figure}

%
For the first fault localization task, we ask the LLM to localize the faulty line of code in the faulty program, $P_{F}$.
It correctly identifies the faulty line of code,  as shown in Figure\ref{fig:LLM_debug_motiv}-Middle.
To measure the robustness of LLM, we create the second fault localization task by asking LLMs to find the faulty line of code in the program, $P_{F+SPM}$. 
An in-depth, high-quality code reasoning would allow LLM to discard any changes that do not impact the code semantics while continuing to find the fault it identified earlier.   
However, as shown in Figure~\ref{fig:LLM_debug_motiv}, LLM does not identify the faulty line of code in the presence of these SPMs. 
Instead, it erroneously flags line 19, \texttt{result.append(list())}, as problematic, while the off-by-one error in \texttt{is\_safe} function remains undetected.
While Claude Sonnet 4.5 correctly detects the fault with respect to the program’s specification, its lack of robustness causes its code reasoning to be easily influenced by semantically irrelevant changes.

%% file: section/research.tex

\section{Research Questions}
To systematically evaluate the accuracy, robustness, and evolution of LLMs' fault localizability,  we investigate the following research questions.

\begin{enumerate}[leftmargin=*]

\item \textbf{RQ1: Robustness of LLMs to Semantic-Preserving Mutations}: 
How robust are LLMs’ fault localization abilities when programs are subjected to SPMs?

\item \textbf{RQ2: Effect of SPM Types and Strengths on Fault Localization}: 
How do different types and strengths of SPMs affect LLMs’ fault localization performance?

\item \textbf{RQ3: Effect of Fault Location on LLMs’ Fault Localization Ability}: 
How does the location of a fault within a program influence LLMs’ ability to localize faults?

\item \textbf{RQ4: Differences Across LLM Categories}: 
How do different categories of LLMs differ in fault localization performance under SPMs?

\item \textbf{RQ5: Longitudinal Trends in LLM Fault Localization}: 
Do LLMs exhibit measurable improvements in fault localization robustness over time as models evolve?

\end{enumerate}

%% file: section/Methodology.tex
\begin{figure*}[t]
  \centering
  \includegraphics[width=\textwidth,keepaspectratio]{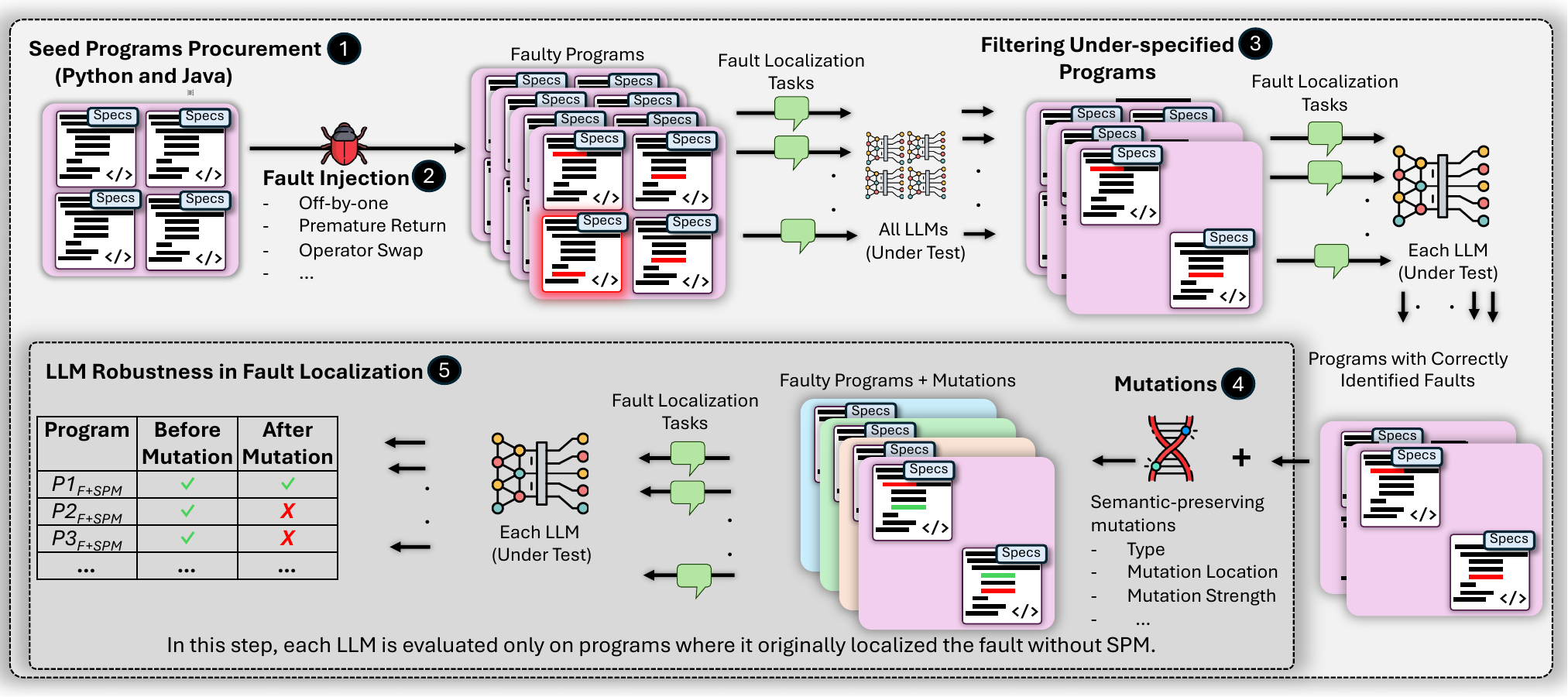}
  \caption{Overview of our methodology, which consists of seed program procurement~\ding{202}, controlled fault injection~\ding{203}, filtering of under-specified programs via fault localization~\ding{204}, and robustness evaluation using semantic preserving mutations~\ding{205} --~\ding{206}.}
  \label{fig:overview}
  \vspace{-2ex}
\end{figure*}

\section{Methodology}

The goal of this work is to dynamically generate an arbitrarily large number of fault localization tasks, with labeled, correct responses expected from LLMs, to test their accuracy and robustness in code reasoning for fault localization. To that end, we design an automated end-to-end evaluation framework to measure an LLM’s fault localization ability, as shown in Figure~\ref{fig:overview}. In this section, we explain the design of this evaluation framework.
%

%

%
Given a set of seed programs \ding{202}, we automatically inject faults into the programs \ding{203} and ask LLMs to localize the faults \ding{204}.
We then exclude fault-localization tasks with underspecified programs, since failure to localize a fault in such programs is due to insufficient specification. Next, we evaluate LLMs on the remaining fault-injected programs by recording their fault localization performance.
We then take the programs that an LLM previously localized correctly and automatically inject semantic-preserving mutations \ding{205} with varying strengths, types, and locations, thereby creating a large number of unique fault localization programs for which ground truth is available.
Finally, we send these mutated fault localization programs to LLMs and record their responses \ding{206}, capturing fault localization accuracy and identifying LLM robustness in the presence of non-functional code changes.

\newcommand{\mc}{\text{\footnotesize $\mathcal{M}_{c}$}}
\newcommand{\mv}{\text{\footnotesize $\mathcal{M}_{v}$}}
\newcommand{\md}{\text{\footnotesize $\mathcal{M}_{d}$}}
\newcommand{\mf}{\text{\footnotesize $\mathcal{M}_{f}$}}
\newcommand{\p}{\text{\footnotesize $\mathcal{P}$}}

\subsection{Seed Programs Procurement}
We focus this empirical study on the two dominant languages, Python and Java, used in code generation benchmarks~\cite{BugInHayStack}. 
Both languages are widely used in open-source projects~\cite{popularLangs, twist2025studyllmspreferenceslibraries}, consequently providing adequate training opportunities for LLMs. 
%
Evaluating LLMs on these languages provides the best opportunity to demonstrate their performance, compared to languages with limited training data.

\begin{figure}[t]
    \vspace{-2ex}
    \centering
    \includegraphics[width=0.9\columnwidth,keepaspectratio]{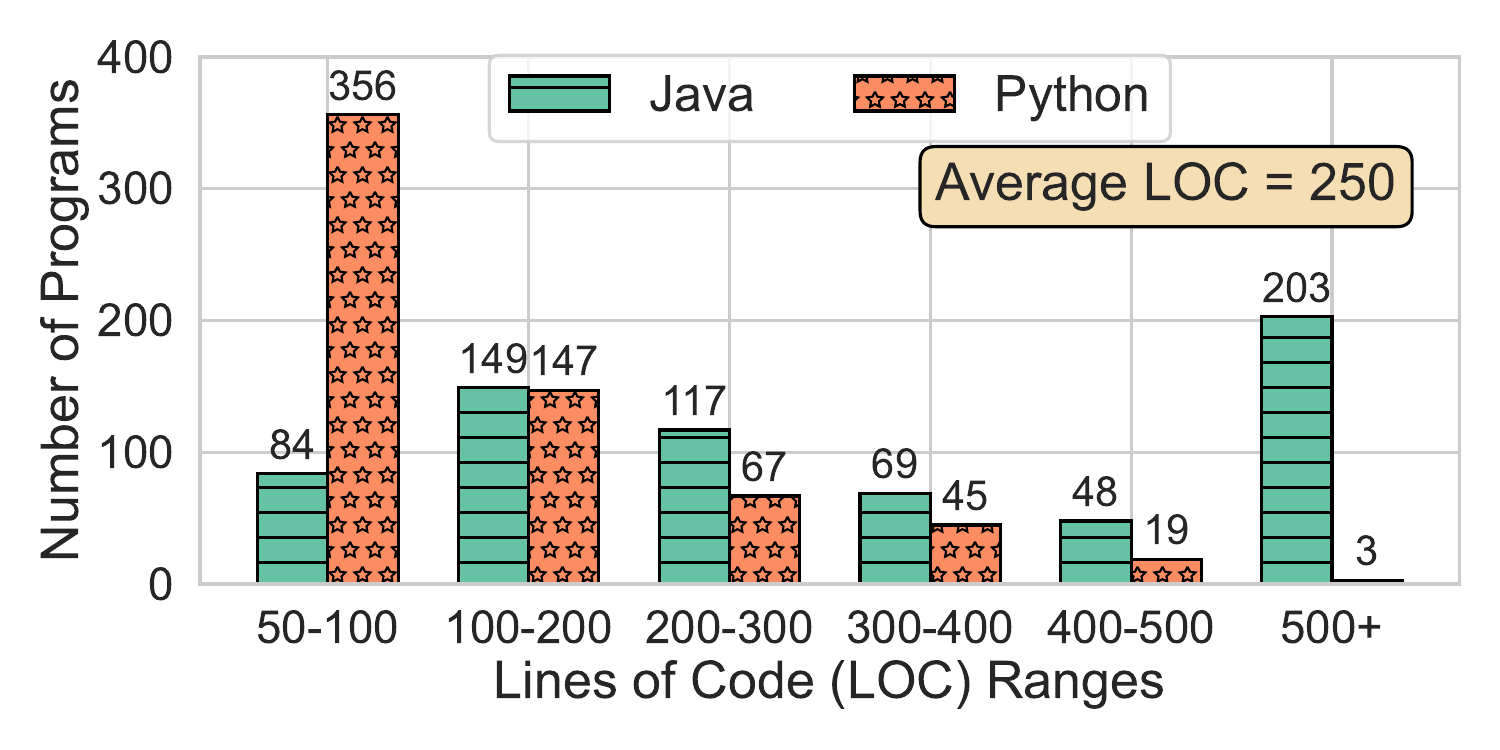}
    \caption{LOC Distribution for the Final Set of Programs}
    \label{fig:program_distribution}
        \vspace{-2ex}
\end{figure}
 
\noindent\textbf{Dataset criteria.}
We use the following criteria to identify the seed programs. 
First, the programs must be accompanied by natural language specifications.
Without specifying the expected semantics of the code, the notion of correctness/incorrectness for fault localization will remain unclear. 
Second, the program must have at least 50 lines of code. LLM benchmarks containing small ($<50$ LOC) toy programs do not represent real-world programs~\cite{LLMToyPrograms}. 
%
Third, we avoid using already faulty programs or fault benchmarks  (e.g., Defects4J~\cite{10.1145/2610384.2628055} or BugsInPy~\cite{10.1145/3368089.3417943}), as prior work~\cite{mem-bug,deng-etal-2024-investigating,dong-etal-2024-generalization} suggests that these datasets have been seen during LLM training, leading to data contamination. 
Lastly, the complete size of each program should not exceed the prompt size limits of the current LLMs APIs.
%
Our focus is not to perform adversarial attacks on LLMs by creating challenging prompt scenarios.
Instead, we aim to identify systematic patterns in code characteristics that reveal weaknesses in LLM fault localization ability.
\begin{table*}[t]
  \scriptsize
  \centering
  \renewcommand{\arraystretch}{1.0}
  \setlength{\tabcolsep}{4pt}
  \begin{tabular}{p{0.18\textwidth} p{0.12\textwidth} p{0.43\textwidth} p{0.2\textwidth}}
    \noalign{\hrule height 1pt}
    \rowcolor{pink!25}
    \textbf{Modification} & \textbf{Type} & \textbf{Description} & \textbf{Example} \\
    \noalign{\hrule height 1pt}
    Off-By-One             & Fault-Inducing      & Alters the loop range, causing an off-by-one error.                  & \texttt{for i in range(n)} → \texttt{for i in range(n+1)} \\
    Misplaced Return       & Fault-Inducing      & Adds a return statement at an unintended location, leading to early termination. & 
    {\tt a=4; b=2+a;} → {\tt a=4; return; b=2+a;} \\
    Boolean Logic          & Fault-Inducing      & Switches boolean operators                                           & \texttt{a \&\& b} → \texttt{a || b} \\
    Operator Swap          & Fault-Inducing      & Switches arithmetic operators                                        & \texttt{a + b} → \texttt{a - b} \\
    Dead Code Injection (\md)   & Semantic-Preserving & Adds code that does not execute or is unused. Increases complexity without changing semantics. & \texttt{if(False): x = 5} \\
    Misleading Comments (\mc)    & Semantic-Preserving & Replaces comments with misleading but coherent descriptions.        & \texttt{/* Summon ancient dragons */} \\
    Misleading Variable Names (\mv) & Semantic-Preserving & Replaces variable names with ones that obscure their real function. & \texttt{count} → \texttt{index} \\
    Function Shuffling (\mf)     & Semantic-Preserving & Shuffles the order of function definitions without breaking dependencies. Only on Java. & \texttt{void fA()\{\}; void fB()\{\}} → \texttt{void fB()\{\}; void fA()\{\}} \\
    \bottomrule
  \end{tabular}
  \caption{Types of Faults and Mutations Applied to Seed Programs}
  \label{tab:mutations}
\end{table*}

\noindent\textbf{Dataset Selection.}
We find two public benchmarks of Python and Java that satisfy the criteria above. We obtain Python programs from~\cite{iamtarun_python_2023} and Java programs from CodeSearchNet~\cite{AhmedSSoliman_CodeSearchNet}. 
The dataset consists of $18612$ Python programs and $812$ Java programs comparable to top LLM benchmarks~\cite{cruxeval,ESESizeRef}.
%
After applying the size and context limit filter, we get the final set of 637 Python and 670 Java programs. 
Figure \ref{fig:program_distribution} presents the line-of-code (LOC) distribution of these programs.


\subsection{Fault Injection and Localizability}
We adopt faults from standard mutation testing research~\cite{major}. 
We focus on (1) faults that are confined to a single line, (2) faults pertaining solely to the logical elements of the top-level code and cannot stem from dependencies, and (3) faults that should affect the program’s logic rather than introduce syntax errors. 
Since LLMs today process code sequentially, token-by-token, {\em their understanding of code may vary across different program locations}. 
To discover further evidence, we randomly selected program locations from the  0-25\%, 25\%-50\%, 50\%-75\%, and 75\%-100\% lines of code to inject fault. 
Table~\ref{tab:mutations} presents the types of injected mutations listed under the {\tt Type} {\tt fault-inducing} mutations. 
For instance, incorrect boolean logic and arithmetic operators swap are LOR and AOR mutation operators from MAJOR~\cite{major}, whereas premature return and off-by-one are from ~\cite{offByOne,bugTypes}. 
Our rationale for using simple, standard mutations is to avoid challenging, rare cases that intentionally conceal faults and to provide a fair evaluation setting. If LLMs struggle to localize faults under these standard mutations, they are likely to struggle even more under more challenging faults.

\input{images/LLM_accuracy_python_java}

\noindent\textbf{Filtering Under-specified Fault Localization Tasks.} Before assessing LLMs' faulty localizability, we must ensure that the programs are not under-specified, as correct fault localization for such programs is ambiguous. We treat specification inadequacy as a falsifiable hypothesis and apply a counterexample-based existential criterion: if one or more LLMs localize the injected fault successfully, we retain that FL task as solvable under the given specification. This is shown in \ding{204} in Figure~\ref{fig:overview}.

\noindent\textbf{Fault Localizability.} After fault injection and excluding under-specified FL tasks, we provide a set of ten LLM models with a prompt that includes the program, its natural-language specification, and the task. 
The ten LLMs, listed in Table~\ref{tab:llms}, are selected based on their popularity and include widely used proprietary and open-source alternatives.  Open-source models were executed on local server machines with NVIDIA L40S GPU. Meanwhile, closed-source models were accessed via their respective APIs. 
Once the LLM identifies the line at fault, we automatically compare it with the recorded line number from the fault injection. 
%
We report and analyze the fault localization accuracy of these models on these FL tasks in Section~\ref{LLM-Accuracy-After-Mutation}.  
If an LLM correctly identifies the fault, only then the faulty program is passed to the next step, where robustness is evaluated on the same LLM. 

\begin{table}[t]
  \centering
  \scriptsize
  \setlength{\tabcolsep}{6pt}
  \renewcommand{\arraystretch}{1.0}
  \begin{tabular}{l c c}
    \toprule
    \rowcolor{pink!25}
    \textbf{Model} & \textbf{Size} & \textbf{Type} \\
    \midrule

    Qwen2.5-coder \cite{hui2024qwen25codertechnicalreport} & 7 B  & Open-source \\
    Llama3.1 \cite{grattafiori2024llama3herdmodels}       & 8 B  & Open-source \\
    Phi4 \cite{abdin2024phi4technicalreport}              & 14 B & Open-source \\
    Qwen-QWQ \cite{qwq32b}                                & 32 B & Open-source \\
    GPT-4o \cite{openai2024gpt4ocard}                     & Undisclosed    & Closed-source \\
    Claude 3.7 Sonnet \cite{claude37}                     & Undisclosed    & Closed-source \\
    Gemini 2.0-Flash \cite{gemini2.0flash}                & Undisclosed    & Closed-source \\
    Gemini 1.5-Pro \cite{gemini15pro}                     & Undisclosed    & Closed-source \\
    Gemini 2.5-Flash \cite{gemini25flash}                     & Undisclosed    & Closed-source \\
    Claude 4.5-Sonnet \cite{claude45sonnet}                     & Undisclosed    & Closed-source \\
    \bottomrule
  \end{tabular}
  \caption{LLMs Evaluated}
  \label{tab:llms}
\end{table}

\subsection{LLM's Robustness in Fault Localization}
This step evaluates the robustness of the LLM's code reasoning by applying semantic-preserving mutations to faulty programs on which it previously localized faults correctly. 
Our insight is that if an LLM can adequately reason about code, it should be able to ignore semantically irrelevant changes that do not affect program functionality, thereby maintaining its fault-localization performance.  
The benefits of this process are threefold. 
First, these semantic-preserving mutations provide a stepwise dial to evaluate the limits of the LLM without manually writing new specifications, which can be infeasible at scale. 
Second, semantic-preserving mutations reaffirm that the LLM under test is not just syntactically comparing the program  (i.e., a form of shallow reasoning) with a version it has seen in its training data. 
Lastly, evaluating robustness with the same LLM that solved the FL task confirms the specification is sufficient for the model to reason about the program and that the task is indeed solvable.


On programs that an LLM has successfully localized fault before, we apply semantic preserving mutations of different characteristics (\ding{205} in Figure~\ref{fig:overview}), prepare a fault localization task, and ask the same LLM to localize the same fault again without any context of the previous fault localization task (\ding{206} in Figure~\ref{fig:overview}). 
%
We develop four categories of such mutations:
\begin{itemize}[topsep=0pt,leftmargin=*]
    \item \textbf{Annotative:} Changes to non-executing parts of the code, such as comments, annotations, or metadata. They help evaluate how much LLMs rely on annotations or documentation for reasoning.
    \item \textbf{Identifier:} Changes to the names of variables, functions, or other identifiers. This tests the resilience of LLMs and whether their reasoning is tied to concrete code structure rather than abstract. 
    \item \textbf{Structural:} Changes to the program structure that do not modify functionality. For example, inserting unreachable statements. This helps assess if LLMs ignore semantically irrelevant code. 
    \item \textbf{Non-additive:} Changes to the code order without introducing new content or changing semantics. Examples include changing the order of function declarations.
\end{itemize}

For these four categories, we develop one representative mutation for each, summarized in Table \ref{tab:mutations} under {\tt Type} ``Semantic-Preserving'' category. 
For Dead Code Injection, Misleading Comments, and Misleading Variable Names, the core elements, e.g., the content of the comments, the names of the variables, and the snippets of dead code, are generated via the LLM under test. 
We insert these components via AST manipulation. 
During the insertions, we continuously track the movement of the original fault,  allowing us to still reliably identify faulty lines, even if their position shifts due to formatting changes. 
Similar to fault injection, our rationale for using simple semantic-preserving mutations is to avoid challenging, rare cases that intentionally conceal faults or distort LLMs' reasoning under unreasonably high noise volume. 
If LLMs struggle to localize faults under simple SPMs, they will likely perform even worse under larger, more complex ones.

\noindent\textbf{Mutation applications.}  
We generate multiple program variants by applying mutations both individually and in combination. 
This means that given a program \p, we construct a total of six program mutants. 
Using the notation in Table~\ref{tab:mutations}, four mutants are obtained by applying the individual mutations: \mc(\p), \mv(\p), \mv(\p), \mf(\p); and two more are obtained by composing: \mc(\mv(\p)) and \mc(\mv(\md(\p))). 
For each mutation application, we randomly select program location from 0-25\%, 25\%-50\%, 50\%-75\%, and 75\%-100\% percentiles of the lines of code to apply the mutations. 
We also change the strength of the mutation, e.g., the number of lines of dead code. 
By layering mutations incrementally, we create a structured progression of fault localization difficulty. 

\noindent\textbf{LLMs robustness.} 
The resulting fault localization tasks are sent to the LLM that correctly localized the fault in the previous step.
%
Once a model returns the predicted line number, it is compared against the expected faulty line position.
The robustness accuracies of the models are reported in Section~\ref{LLM-Accuracy-After-Mutation}.

\subsection{Evaluation Framework Implementation}
We develop this evaluation framework in Python.
Seed programs are provided as JSON files containing a natural-language specification and the corresponding source code. 
Both fault injection and semantic-preserving mutation can be provided as AST transformation scripts using existing templates. 

Both proprietary LLMs' APIs and local models are supported. Local inference is enabled via the Ollama runtime, allowing models to be specified without code changes. The framework automatically manages prompt construction, structured response parsing, and result aggregation.

%% file: images/LLM_accuracy_python_java.tex
\setlength{\tabcolsep}{2pt} 
\renewcommand{\arraystretch}{1.15} 
\begin{table*}[t]
\centering
\scriptsize
\begin{tabularx}{\linewidth}{l|*{10}{>{\centering\arraybackslash}X}}
    \noalign{\hrule height 1pt}
    \rowcolor{pink!25}
    \textbf{Fault Type} &
    \textbf{Gemini 2.0 Flash} &
    \textbf{Gemini 1.5 Pro} &
    \textbf{GPT-4o} &
    \textbf{Llama 3.1} &
    \textbf{Phi-4} &
    \textbf{Qwen 2.5-coder} &
    \textbf{Qwen-QwQ} &
    \textbf{Claude-3.7-Sonnet} &
    \textbf{Gemini-2.5-Flash} &
    \textbf{Claude-4.5-Sonnet} \\
    \noalign{\hrule height 1pt}

    \rowcolor{pink!25}
    \multicolumn{11}{c}{\textbf{Python Benchmark Programs}} \\
    \noalign{\hrule height 1pt}

    IncorrectBooleanLogic & 25.22 & 32.74 & 20.35 & 11.95 & 19.91 & 13.27 & 15.49 & \textbf{36.28} & 27.88 & 31.86 \\
    MisplacedReturn      & 40.08 & 29.60 & 23.25 & 9.31 & 23.96 & 11.14 & 14.29 & 44.76 & 35.40 & \textbf{58.75} \\
    OffByOne             & 30.13 & 31.45 & 32.65 & 16.33 & 19.45 & 17.53 & 16.69 & 33.49 & 30.61 & \textbf{44.78} \\
    OperatorSwap         & 29.07 & \textbf{37.41} & 28.12 & 15.81 & 22.25 & 12.41 & 14.02 & 36.55 & 35.13 & 32.86 \\

    \noalign{\hrule height 1pt}
    \rowcolor{pink!25}
    \multicolumn{11}{c}{\textbf{Java Benchmark Programs}} \\
    \noalign{\hrule height 1pt}

    IncorrectBooleanLogic & 14.43 & 25.00 & 17.27 & 6.44 & 10.57 & 9.79 & 11.08 & 51.80 & 17.27 & \textbf{71.13} \\
    MisplacedReturn      & 36.37 & 39.63 & 23.34 & 6.87 & 15.77 & 8.50 & 15.77 & 11.16 & 31.20 & \textbf{41.67} \\
    OffByOne             & 37.50 & 33.33 & 16.67 & 8.33 & 8.33 & 4.17 & 4.17 & 70.83 & 16.67 & \textbf{83.33} \\
    OperatorSwap         & 14.50 & 28.73 & 27.91 & 10.81 & 13.27 & 4.24 & 19.43 & \textbf{68.95} & 20.79 & 31.74 \\

    \noalign{\hrule height 1pt}
\end{tabularx}
\caption{Detection accuracy of different LLMs on various bug types in Python and Java (computed from valid programs).}
\label{tab:bug_detection_csv}
\end{table*}

%% file: section/Evaluation.tex
\section{Results} \label{sec:evaluation}
We analyze the performance of the LLMs across various dimensions, including fault type, fault location, language, SPM type, SPM strength, SPM location, and LLM category. We also conduct a longitudinal study that tracks performance evolution across different versions of the same state-of-the-art models. We create a total of over 750K debugging tasks for these LLMs, spanning 245 million lines of code,  and 3.8 billion tokens. Table~\ref{tab:experimental_setup} documents the experiment statistics. 
\setlength{\aboverulesep}{0pt}
\setlength{\belowrulesep}{0pt}

\begin{table}[ht]
    \centering
    \scriptsize
    \begin{tabular}{l r}
        \toprule
        \rowcolor{pink!25}
        \textbf{Metric} & \textbf{Value} \\
        \midrule
        Total Prompts Generated & 1,163,686 \\             
        Total FL Tasks   & 750,013 \\
        Average LOC  Tested           & 250 LOC\\
        Total LOC Tested          & 245,856,641 LOC \\
        Total Token Analyzed      & 3,881,577,500 \\
        System Specification      & 48 GB RAM, 48 cores with NVIDIA L40S GPU \\
        APIs      & Gemini, Anthropic, OpenAI \\
        \bottomrule
    \end{tabular}
    \caption{Experimental Statistics }
    \label{tab:experimental_setup}
    \vspace{-3ex}
\end{table}

\begin{figure*}[ht]
  \centering
  \includegraphics[width=\textwidth]{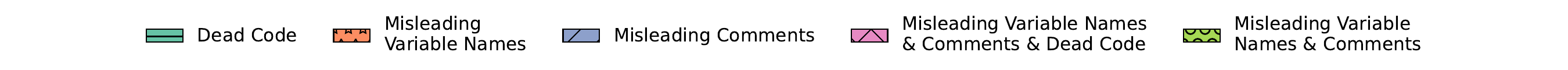}
  \includegraphics[width=\textwidth]{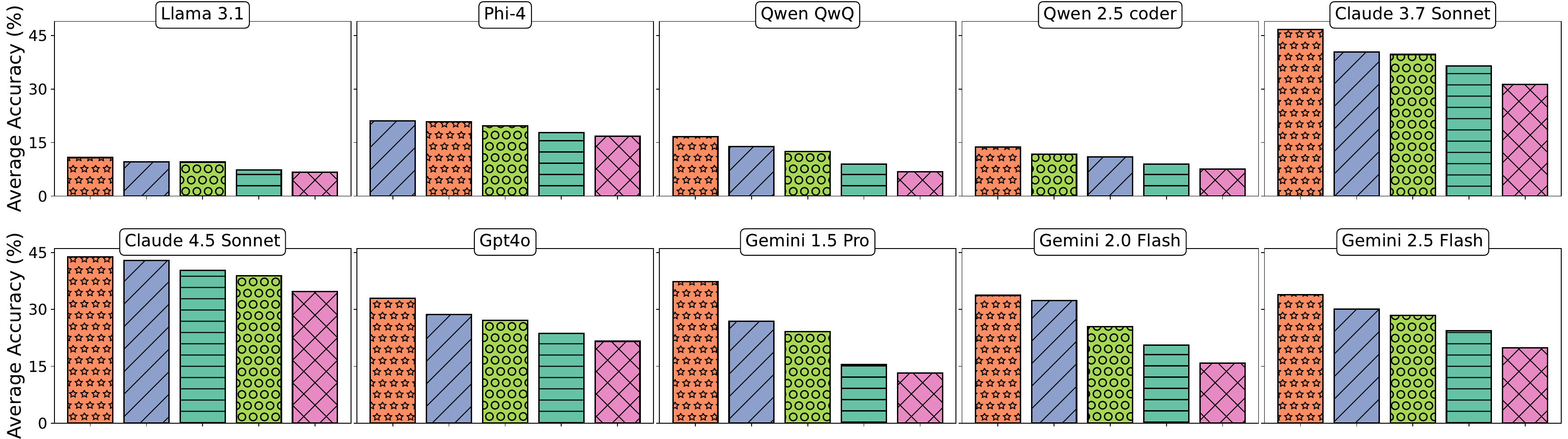}
  \caption{Effect of mutation type on fault detection accuracy averaged for Python~\cite{iamtarun_python_2023} and Java datasets~\cite{AhmedSSoliman_CodeSearchNet}.}
  \label{fig:mutation_type_wise_results}
\end{figure*}

\subsection{RQ1: Robustness of LLMs' Fault Localization}
\label{LLM-Accuracy-After-Mutation}
First, we analyze the fault localization accuracy of individual LLMs. This is to establish a baseline for observing how performance degrades as we test the robustness of LLMs via SPMs. We exclude under-specified FL tasks.  Table~\ref{tab:bug_detection_csv} presents these results. Accuracy is the ratio of faulty programs for which the LLM successfully localizes faults to the total number of faulty programs that are not underspecified. 

The Claude family of models achieves the best overall performance in fault localization across both languages. Except Claude, almost all models show better performance in Python than in Java. Certain faults are better detected by LLMs than others. For example,  7 out of 10 models show the best performance on \textit{MisplacedReturn}. This is consistent with token-sequential processing by LLMs.  A return statement changes which later tokens are interpreted as reachable, shifting attention and, in turn, the model’s fault localization reasoning. 

\textit{Open-Source vs. Closed-Source LLMs.} Notably, Claude 4.5 Sonnet, Claude 3.7 Sonnet, and Gemini 1.5 Pro lead in fault detection performance, demonstrating higher accuracy of fault localization, followed closely by OpenAI’s GPT‑4o. The highest fault detection accuracies come from closed‑source models, suggesting that proprietary LLMs benefit from extensive training data, deployments on hardware needed to host the full‑scale models, and optimizations. Among open‑source LLMs, Phi‑4:14B outperforms other open‑source models. 

\subsubsection{Fault Localization After Applying SPMs}
This analysis measures the impact of semantic-preserving mutations (SPMs) on an LLM's fault localization by asking the LLM to localize the same fault in a mutated faulty program ($P_{F+SPM}$) that it has successfully localized in a faulty program ($P_{F}$).

\begin{figure}[t]
  \centering
  \includegraphics[width=1\columnwidth,keepaspectratio]{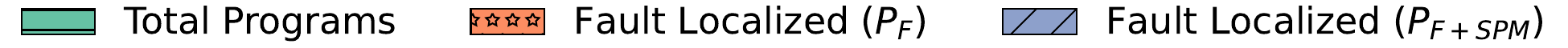}
  \includegraphics[width=1\columnwidth,keepaspectratio]{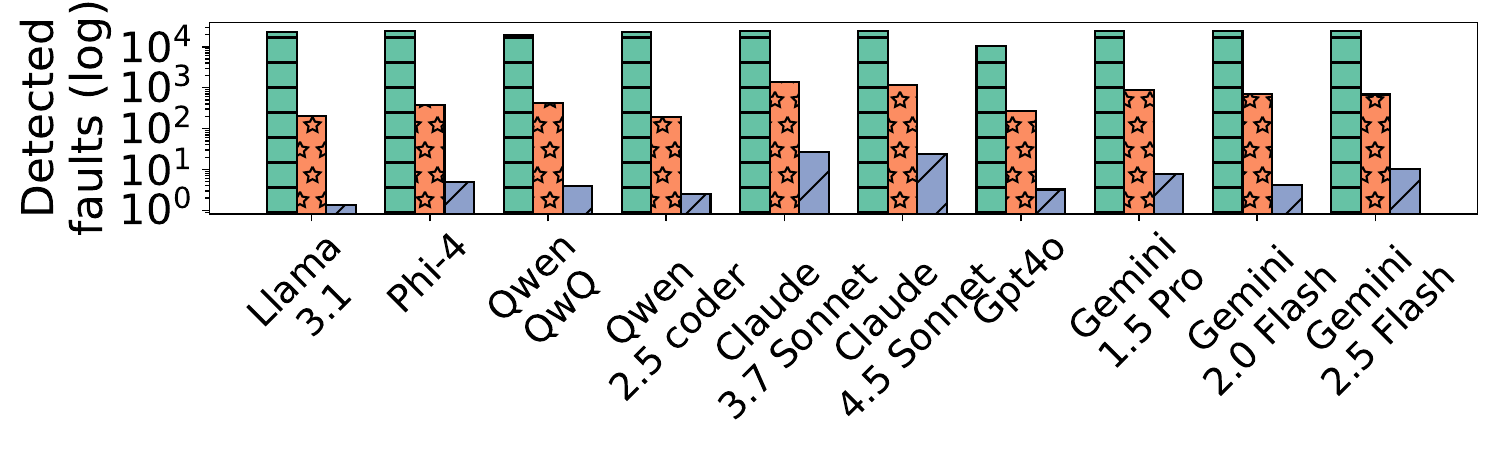}
    \includegraphics[width=1\columnwidth,keepaspectratio]{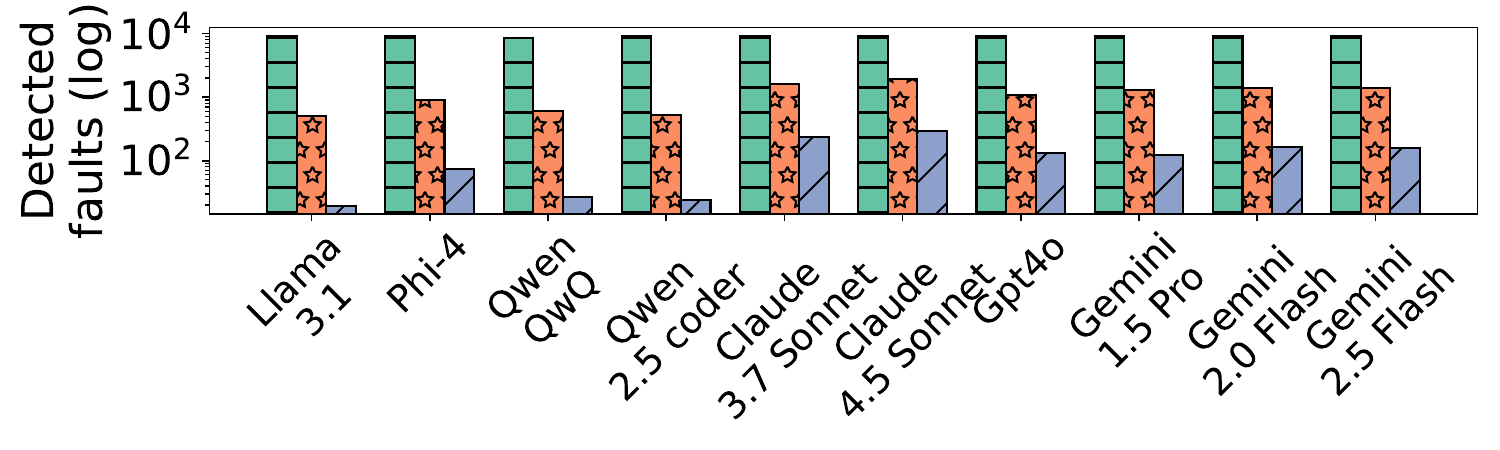}
  \caption{Effect of mutations on fault detection accuracy across all fault types on each LLM with the \underline{Java dataset (top)} and \underline{Python dataset (bottom)}.}  
  \label{fig:mutation_vs_original}
\end{figure}

Figure~\ref{fig:mutation_vs_original} summarizes the results averaged across all fault types and semantics-preserving mutations (SPMs). Accuracy is measured as the ratio of successful fault localization after adding SPMs to the faulty programs, step~\ding{206} of the process. Almost all models exhibit a noticeable degradation in fault localization accuracy when SPMs are introduced. Open-source models, including Llama3.1:8B and Qwen variants, exhibit the most significant performance declines, whereas closed-source models such as Claude, GPT-4o, and Gemini demonstrate greater robustness. 

\begin{tcolorbox}[colback=gray!10!white, colframe=black, boxrule=0.5pt, sharp corners]
\small
\textbf{Takeaway:} LLMs exhibit significant drops in fault localization accuracy under semantic-preserving mutations, revealing difficulty in distinguishing semantically irrelevant code features from semantically relevant ones.
\end{tcolorbox}

\subsubsection{Language-specific Performance under SPMs}
Figure~\ref{fig:mutation_vs_original}  shows the performance breakdown by language.
The results show that LLMs struggle more to detect faults in Java than in Python across all models; with a wider gap for the Claude, GPT-4o, and Gemini models.
This behavior can be attributed to the size and diversity of the training data available for these models, as well as the intrinsic structural differences between Java and Python~\cite{cassano2023multipl}.
Python’s concise syntax and its widespread use in scripting and data science lead to more comprehensive coverage training~\cite{smith2022impact}. 
Conversely, Java’s verbosity and strict object-oriented nature can pose additional challenges, such as maintaining a large context for fault detection. Most models exhibit reduced performance on Java compared with Python. 

\begin{figure*}[t]
  \centering

  \includegraphics[width=0.85\textwidth,trim=6pt 4pt 6pt 2pt,clip]{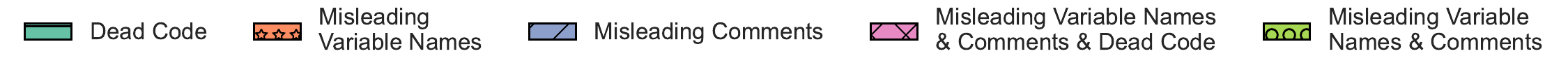}
  \vspace{4pt}

  \begin{subfigure}{\textwidth}
    \centering
    \includegraphics[width=\textwidth,trim=5pt 10pt 12pt 10pt,clip]{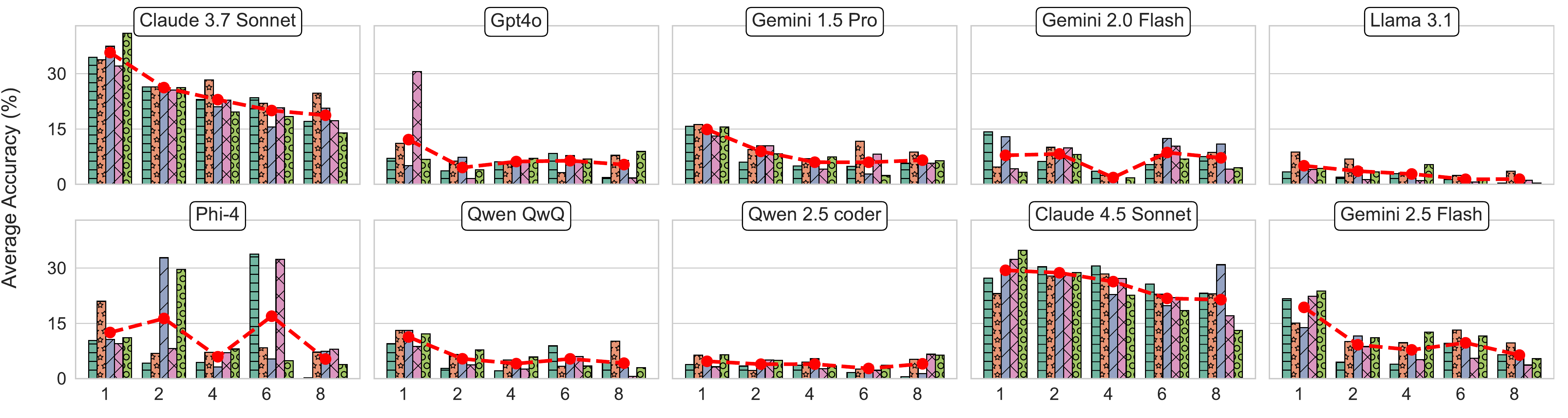}
    \caption{Java programs}
    \label{fig:mutation_strength:a}
  \end{subfigure}

  \vspace{6pt}

  \begin{subfigure}{\textwidth}
    \centering
    \includegraphics[width=\textwidth,trim=5pt 10pt 12pt 10pt,clip]{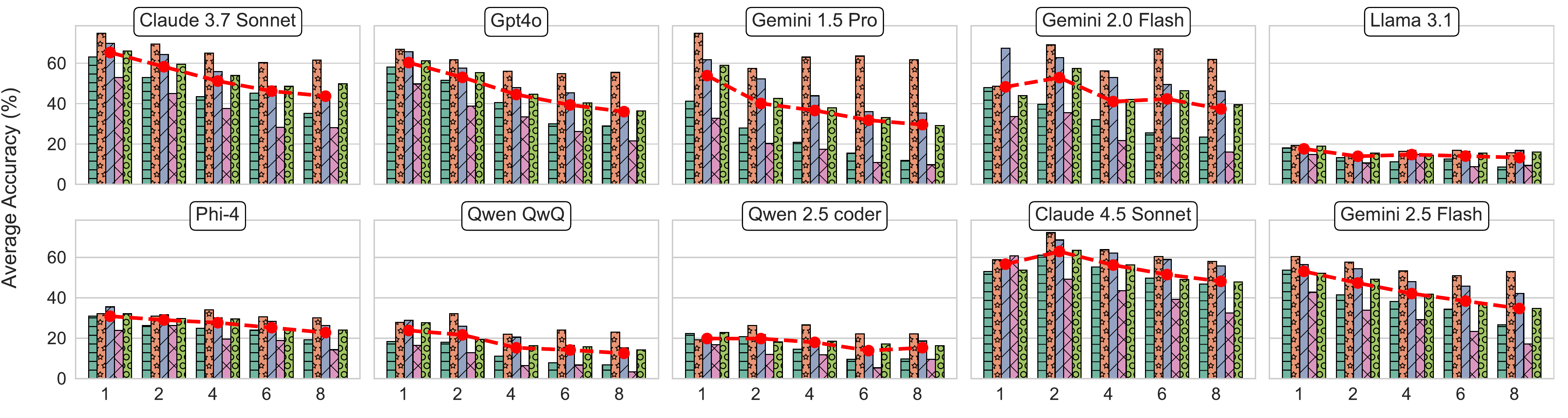}
    \caption{Python programs}
    \label{fig:mutation_strength:b}
  \end{subfigure}

  \caption{Effect of mutation strength on fault detection accuracy by mutation type for different models. Red line shows the average accuracy across mutation strengths.}
  \label{fig:mutation_strength}
  \vspace{-2em}
\end{figure*}

\begin{tcolorbox}[colback=gray!10!white, colframe=black, boxrule=0.5pt, sharp corners]
\small
\textbf{Takeaway:} While LLMs initially achieve higher fault-localization accuracy on Java (without SPMs; Table~\ref{tab:bug_detection_csv}), applying SPMs makes it substantially harder for them to localize the same fault. As a result, the degradation on Java is markedly larger than on Python (Figure~\ref{fig:mutation_vs_original}).
\end{tcolorbox}

\subsection{RQ2: Effect of Mutation Characteristics}

We further decompose results by mutation type and strength.

\subsubsection{Effect of SPM Type on LLM's Fault Localizability}

Figure~\ref{fig:mutation_type_wise_results} presents the average fault localization accuracy across different LLMs for different types of SPMs. \emph{Misleading Variable Names} causes the least impact ($29.02\%$ accuracy) on fault localization accuracy. One reason is that variable renaming preserves the structural and lexical integrity of code. Because LLMs learn statistical patterns from training data that contain variable names of all types, there is no single variable-naming convention; therefore, LLMs are relatively robust to renaming.






\emph{Misleading Comments} result in lower average accuracy ($25.63\%$) compared to \emph{Misleading Variable Names} ($29.02\%$), indicating that LLMs may rely on comments for code reasoning.
\emph{Dead Code} has a more substantial impact on accuracy,  
dropping LLMs' fault localization accuracy drops to $20.38\%$. 
For example, one of the programs simulates an autonomous car; the update method contains the injected fault where the car's vertical movement is adjusted with an incorrect offset (using \texttt{self.rect.y $+=$ self.change\_y $-$ 1}). The program contains an unreachable dead code block that defines a function for debugging sensor values. This dead code changes the syntactic structure without affecting functionality, distracting the model and leading it to erroneously flag the sensor log call within the dead code block as the fault, rather than the movement logic.

\subsubsection{Effect of Function Reordering on Bug Detection Accuracy} Function reordering mutations only apply to Java programs. We observe a significant accuracy drop of $83\%$, demonstrating that even without introducing any new code or logic changes, the ability of LLMs to localize faults is compromised by structural shifts. 
This indicates that the physical location of code plays a prominent role in LLMs' code reasoning.
Because Java is inherently more verbose than Python, these reordering mutations likely push critical logic into deeper parts of the context window where LLMs are known to exhibit reduced reasoning capacity~\cite{du2025context, liu2024lost}.

\subsubsection{Effect of Mutation Strength}
\label{Mutation Strength Effect}
Figure~\ref{fig:mutation_strength} presents the results for each mutation type with varying strength levels (1 to 8), where strength represents the number of times an SPM was applied within a single program. For most LLMs, we observe a linear decline in average fault localization accuracy (indicated by the red line) as SPM strength increases. 
For instance, in Java, the accuracy drops on average from $15.82\%$ at strength 1 to $8.57\%$ at strength 8 across all evaluated models, corresponding to an overall decrease of $7.25\%$ (an average decrease of $1.04\%$ per mutation strength increase). Similarly, in Python, the accuracy falls, on average, from $42.88\%$ at strength 1 to $29.34\%$, which represents an overall decline of $13.54\%$ over 7 steps (an average decrease of $1.93\%$ per mutation strength increase).  In one seed Python program, the Blender API function \texttt{make\_tile} is injected  with an early \texttt{return None} fault. Gemini 1.5 Pro localizes the fault with low dead-code SPM strength, but fails when dead-code strength increases to 4 (multiple dead statements).

\begin{tcolorbox}[colback=gray!10!white, colframe=black, boxrule=0.5pt, sharp corners]
\small
\textbf{Takeaway:} Misleading comments and dead code cause the highest disruption in LLMs fault localizability, and even minor code reordering (e.g., function reordering) can reduce fault localization accuracy, highlighting LLMs’ reliance on surface-level code cues.
\end{tcolorbox}


\subsection{RQ3: Effect of Fault Location in the Code}



\begin{wrapfigure}[12]{r}{0.6\columnwidth}
    \centering
    \captionsetup{skip=2pt} 
    \includegraphics[width=\linewidth, height=0.35\textheight, keepaspectratio]{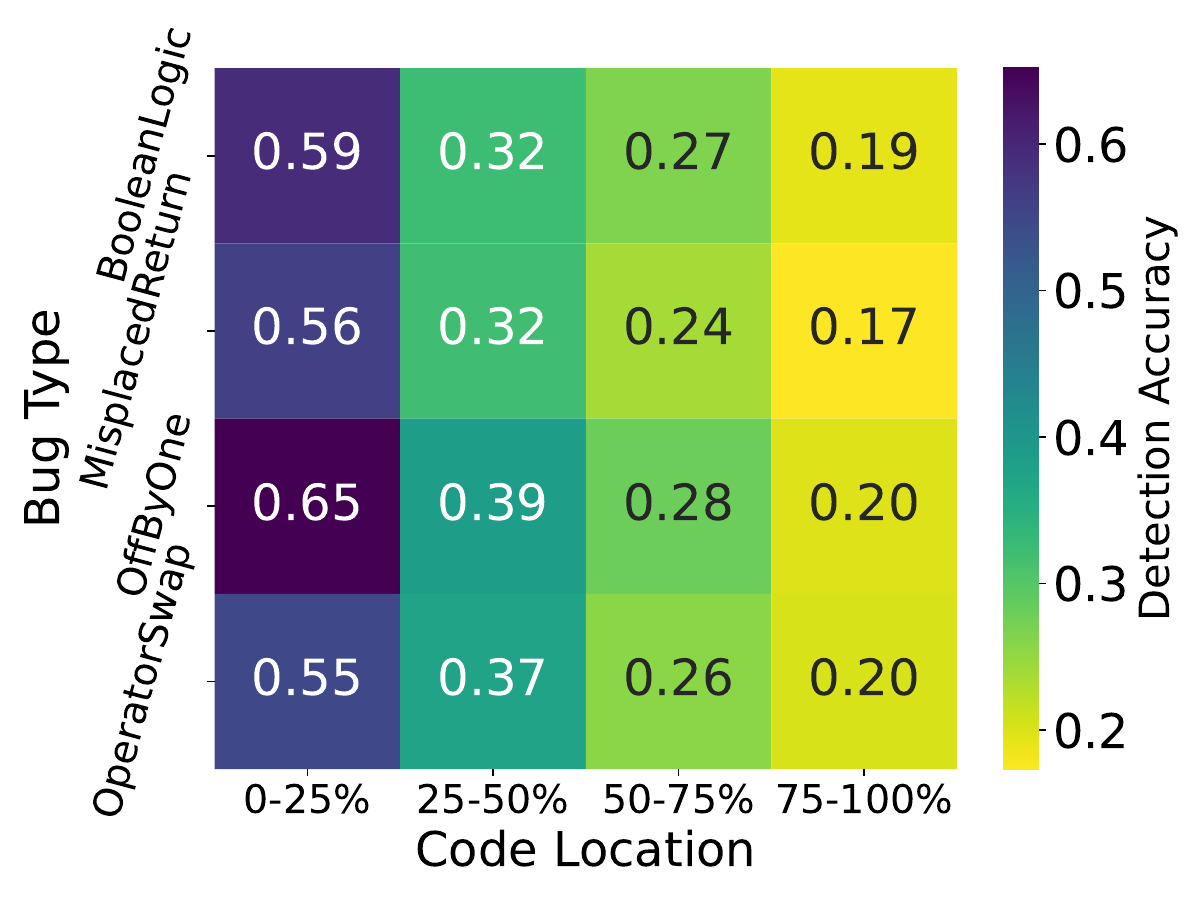}
    \caption{Effect of fault location on fault detection accuracy.}
    \label{fig:heatmap_loc}
\end{wrapfigure}

Figure \ref{fig:heatmap_loc} presents a heatmap summarizing fault localization accuracy across different fault types and their positions within the code. Results show that faults in the first quarter of the code ($0-25\%$) are detected with the highest accuracy, suggesting that LLMs may focus more on initial segments and retain clearer context at the beginning of a program. 
\setlength{\intextsep}{1pt}  
\setlength{\columnsep}{5pt}  
\noindent
In particular, {\it OffByOne} faults are easily detected in this early section. However, detection accuracy declines as faults appear later in the code, with the $75-100\%$ range exhibiting the lowest success rates across all fault types. 

These findings suggest that LLMs may lose context or allocate less attention to code segments appearing farther from the start due to the cumulative noise and attention decay inherent in long-sequence processing \cite{hong2025contextrot}. 

\begin{tcolorbox}[colback=gray!10!white, colframe=black, boxrule=0.5pt, sharp corners]
\small
\textbf{Takeaway:} LLMs are better at localizing faults in the early code regions, with accuracy declining for faults in later sections, indicating limited context retention and positional non-linear reasoning ability across code regions.
\end{tcolorbox}
\setlength{\intextsep}{-1em}  


\subsection{RQ4: Categories of LLMs}
\begin{wrapfigure}[9]{r}{0.5\columnwidth}
    \vspace{0.7em}
    \centering
    \captionsetup{skip=2pt} 
    \includegraphics[width=\linewidth, height=0.35\textheight, keepaspectratio]{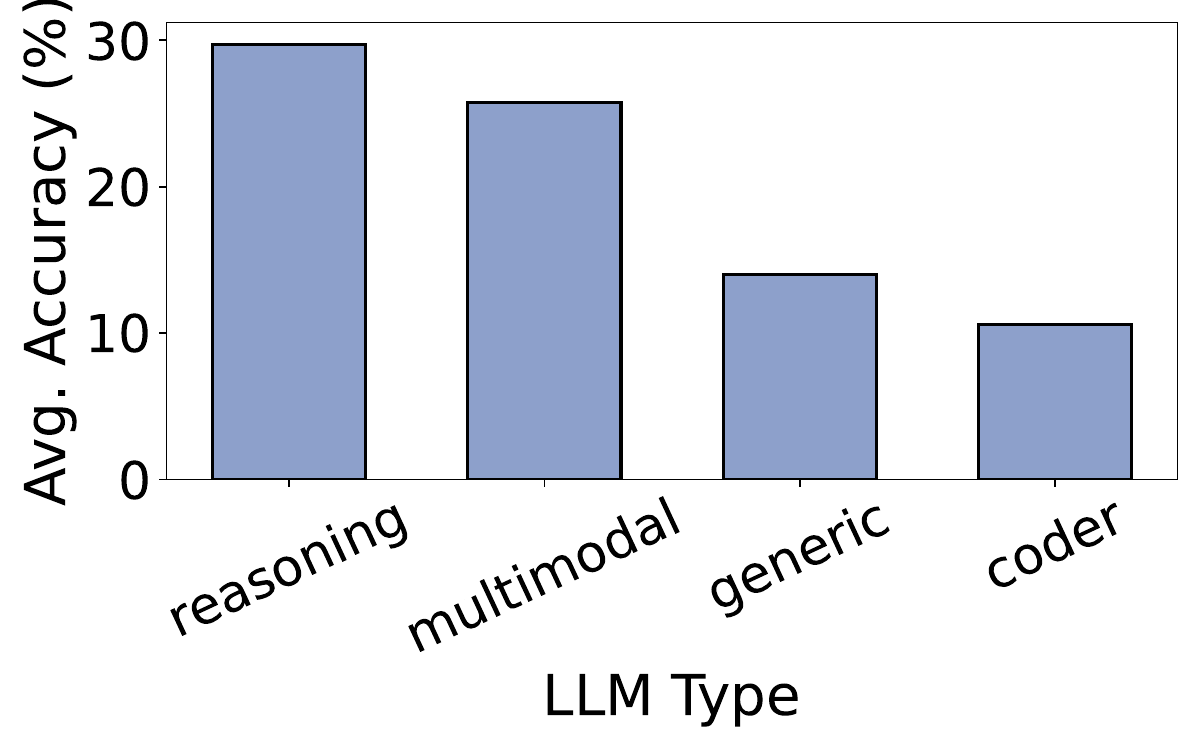}
    \caption{Effect of LLM Type on fault detection accuracy.}
    \label{fig:Llm_type_accuracy}
\end{wrapfigure}
Figure \ref{fig:Llm_type_accuracy} shows that reasoning and multimodal models like Claude and Gemini achieve the highest accuracy, while general-purpose models like Llama 3.1 and Phi-4, and coding-specialized models like Qwen2.5-coder:7B, yield the lowest performance. This discrepancy highlights a broader insight: LLMs optimized for reasoning may develop more robust internal representations of code semantics due to their exposure to diverse instruction-following and multi-step reasoning tasks, which enhances their ability to look past syntactic noise~\cite{yang-etal-2025-code,han2025beyond,illusion-of-thinking}.

\subsection{RQ5: Longitudinal Study of LLMs}
\label{subsec: Longitudinal Study of LLMs}

\begin{figure}
    \centering
    \captionsetup{skip=2pt} 
    \includegraphics[width=0.8\linewidth, keepaspectratio]{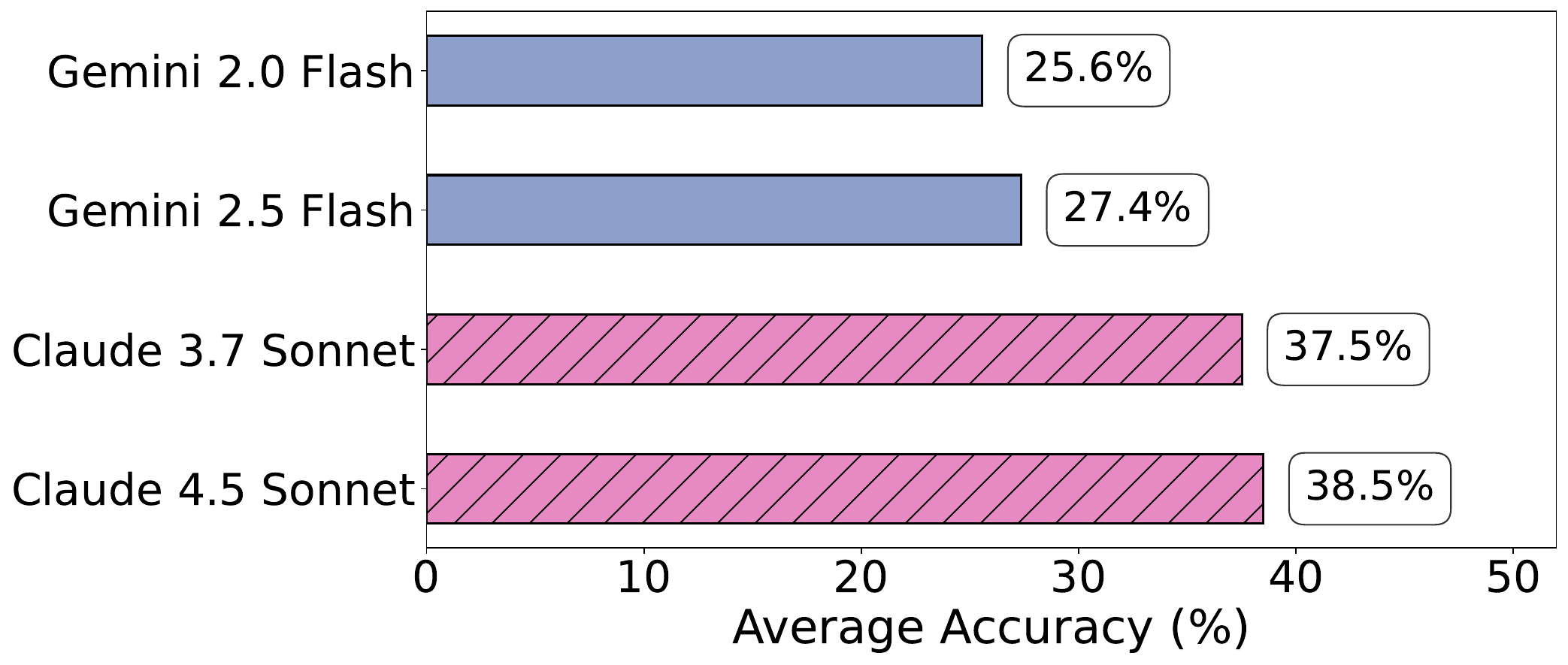}
    \caption{Performance of the same models across versions.}
    \label{fig:gemini_claude_longitudinal_study}
    \vspace{-1.9em}
\end{figure}

We also conduct a longitudinal analysis of the two best-performing model families, Gemini and Claude as shown in Figure \ref{fig:gemini_claude_longitudinal_study}. We observe a $1.8\%$ improvement for Gemini and an average $1.0\%$ improvement across Claude versions. These gains are consistent with expected model iteration effects like additional training data, improved training recipes for coding tasks, and overall capability improvements that translate to better code reasoning for fault localization. While directionally positive, these gains are marginal, underscoring the need for more fundamental advances in how models {\em represent}, {\em interpret}, and {\em prioritize} code semantics to reason more deeply about code logic.

%% file: section/Discussion.tex
\section{Discussions}
In this section, we discuss key findings and evaluation design choices in an FAQ format.




\noindent{\em What is the reason behind selecting single-file programs for such empirical investigation?}~We use simple, single-file programs with clear specifications to give LLMs the most favorable conditions. Even under those best-case settings, models fail to localize faults reliably. Theoretically, as program length grows, attention over many tokens leads to context dilution and positional bias, making it harder to reason about code~\cite{wu2025emergencepositionbiastransformers}. Thus, poor performance on simple programs provides a conservative lower bound.


\noindent{\em What is the rationale for using semantic-preserving mutations?}~By design, classical fault-localization techniques (e.g., delta debugging~\cite{10.1109/32.988498} and mutation-based fault localization~\cite{papadakis2013mbfl} and their extensions) rely on dynamic execution evidence such as tests and coverage, and are therefore robust to non-functional changes by SPMs. As LLMs are increasingly used as substitutes for these techniques in practice, developers expect comparable robustness. We thus argue that LLM-based fault localization should, at a minimum, meet the performance of these classical baselines. We employ simple, semantic-preserving mutations to incrementally increase the difficulty of fault localization tasks in a controlled way. This process is analogous to fuzzing, in which controlled transformations are systematically applied to the input to test the program. 
%

\noindent {\em Can prompt engineering improve LLMs fault localizability?}~ 
We intentionally avoid extensive prompt engineering and use the {\em same} uniform baseline prompt illustrated in Figure~\ref{fig:LLM_debug_motiv}. Although techniques such as interactive multi-shot prompting and iterative refinement might enhance performance, they require significant manual intervention, are unsuitable for large-scale evaluations, and overfit results to a single prompt. 



\noindent {\em What broader implications do our findings have on future directions in code representation?}~
LLMs process code similarly to textual data, relying on generic tokenization methods that overlook code-specific features. We argue that converting code into intermediate, abstract, structured representations (like Control Flow Graphs (CFGs) and Code Property Graph (CPG)~\cite{CPG}) may enhance LLM reasoning capabilities. This approach can better capture syntactic and semantic nuances, improving robustness to non-functional code alterations. 

\section{Threats to Validity.}


Injected faults may not impact all possible execution paths or semantic behaviors. Future work could incorporate program analysis, such as program slicing or path-based fault injection, to improve execution path coverage. The evaluation includes ten LLMs. While this diversity mitigates model-specific bias, the results may not generalize to all existing or future LLMs. Additionally, our fault localization tasks are derived from Python and Java programs, and the findings may not extend to other programming languages. We select four injected fault types from prior mutation-testing literature. This set may not capture the full diversity of real bugs (e.g., multi-location bugs).  To mitigate this, we select faults that are relatively easy to detect to provide a fair evaluation setting. Likewise, our selected SPMs do not cover all semantic-preserving transformations (e.g., refactorings, control-flow restructuring, API-equivalent rewrites). However, the simplicity of our SPMs provides a conservative lower bound on LLM robustness.

%% file: section/related.tex
\section{Related Work}

Table~\ref{tab:comparison_benchmarks} compares major benchmarks across five criteria.
 
\noindent\textbf{Code Generation and Benchmarking.}
LLMs for code are most commonly evaluated on \emph{generation}-centric benchmarks such as HumanEval~\cite{HumanEval} and MBPP~\cite{MBPP}, along with follow-up suites that tighten functional evaluation e.g., HumanEval+~\cite{liu2023is}. However, these benchmarks primarily measure whether a model can \emph{produce} correct code, not whether it can reliably \emph{operate on existing code}\cite{xie2025corebenchmarkingllmscode}.
Recent work has also highlighted that public benchmarks can be contaminated by pretraining or post-release continual training~\cite{deng-etal-2024-investigating,mem-bug,dong-etal-2024-generalization,chen2025dycodeeval}. To mitigate this, “fresh” benchmark constructions such as LiveCodeBench\cite{jain2025livecodebench} and LiveBench\cite{white2025livebench} sample tasks from post-cutoff time windows to reduce overlap with training data, but they still target generation performance rather than maintenance tasks like fault localization~\cite{jain2025livecodebench,white2025livebench}.


\definecolor{DarkGreenCustom}{rgb}{0.0, 0.8, 0.0}
\newcommand{\greencheck}{\Large{{\color{DarkGreenCustom}\checkmark}}}
\newcommand{\redcross}{\large{{\color{red}\ding{55}}}}

\begin{table}[t]
  \centering
  \scriptsize
  \renewcommand{\arraystretch}{1.2}
  \setlength{\tabcolsep}{3.5pt}
  \begin{tabularx}{\columnwidth}{
      >{\raggedright\arraybackslash}X
      c c c c c}
    \toprule
    \textbf{Benchmark} & \textbf{GT} & \textbf{ML}
      & \textbf{CCC} & \textbf{Scal.} & \textbf{RDC} \\
    \midrule
    \textbf{HumanEval+ \cite{liu2023is}} 
      & \greencheck & \redcross & \redcross & \redcross & \redcross \\

    \textbf{DebugBench \cite{tian2024}} 
      & \greencheck & \greencheck & \redcross & \redcross & \redcross \\

    \textbf{LiveCodeBench \cite{jain2025livecodebench}} 
      & \greencheck & \greencheck & \redcross & \redcross & \greencheck \\

    \textbf{SOAPFL \cite{SoapFL}} 
      & \greencheck & \greencheck & \redcross & \redcross & \redcross \\

    \textbf{FlexFL \cite{xu2025flexfl}}
  & \greencheck & \greencheck & \redcross & \greencheck & \redcross \\

    \textbf{Ours} 
      & \greencheck & \greencheck & \greencheck & \greencheck & \greencheck \\
    \bottomrule
  \end{tabularx}

  \caption{Comparison of recent benchmarks under five criteria:
  \textbf{GT}~(Ground Truth),
  \textbf{ML}~(Multi-Language),
  \textbf{CCC}~(Configurable Code Complexity),
  \textbf{Scal.}~(Scalability),
  \textbf{RDC}~(Robust to Data Contamination).}
  \vspace{-0.5em}
  \label{tab:comparison_benchmarks}
\end{table}

\noindent\textbf{Debugging and fault localization.}
Recent research~\cite{Smith2023LLMDebugging, Nguyen2023FaultLocalization, Lee2022DeepDebug,hort2025semanticpreservingtransformationsmutationoperators} utilizes and assesses the debugging capabilities of LLMs. They employ automated fault-localization strategies, including mutation testing and functional test cases, to determine whether an LLM can pinpoint errors; however, on standard fault benchmarks. 
Our work fills this gap by offering a mechanism to evaluate LLM-based FL methods on unseen FL tasks. \cite{ZengDefect2023} perform an evaluation of pre-trained models on code tasks using CodeXGLUE\cite{codexglue} and BigCloneBench\cite{bigclonebench} benchmarks, replicating results on tasks such as code search, generation, and defect detection, but their reliance on fixed datasets risks data contamination from pretraining corpora.  Simiarly, \cite{Liu2025ExeRScope} introduces ExeRScope, a toolkit for in-depth analysis of code execution reasoning frameworks (e.g., CodeMind\cite{codemind}, REval\cite{reval}) on benchmarks like CruxEval\cite{cruxeval}, also depend on static evaluation datasets that may overlap with LLM training data. 
\cite{Risse2024TopScore} examine the internal validity of machine learning for vulnerability detection (ML4VD), which typically involves determining whether a given function is vulnerable without considering broader context. Their work primarily critiques the design of datasets and the framing of problems within ML4VD. In contrast, our work directly interrogates whether LLMs’ \emph{fault localization remains robust} under semantic-preserving code changes.

\noindent\textbf{Program mutation and semantic analysis.}
Program mutation techniques have been used to stress-test model behavior under controlled transformations~\cite{Garcia2023ASTMutation, Johnson2023ASTTransform}. We build on this idea but focus specifically on \emph{fault-localization robustness} under semantic-preserving mutations. Similarly, other works~\cite{Patel2023SemanticAlterations, Wong2024SemanticPreserving, NeurIPS2023LLMStruct, Garcia2023LLMEnhance} have refined these techniques to model the effect of targeted semantic changes. Despite these advances, existing mutation frameworks typically emphasize semantic alterations that impact functionality while neglecting the subtleties of semantic-preserving changes.

%% file: section/conclusion.tex
\section{Conclusion}
While LLMs are increasingly used in software development, their evaluation remains focused on code generation. This paper conducts a large-scale empirical study assessing LLMs' fault localization ability and their robustness under different scenarios. We automatically inject faults and semantic-preserving mutations in existing benchmarks to generate a large amount of unseen debugging tasks for LLMs. Experiments on 750K tasks reveal fundamental weaknesses in LLMs, with non-functional code changes reducing debugging accuracy by 78\%, highlighting high sensitivity and low robustness under semantic-preserving mutations. We identify key code features that challenge LLMs and expose unique weaknesses, guiding research toward more effective LLM use for robust LLM-based debugging systems.




